# Origin of Orbits of Secondaries in the Discovered Trans-Neptunian Binaries

S. I. Ipatov[a, b]

[a]Vernadsky Institute of Geochemistry and Analytical Chemistry, Russian Academy of Sciences, Moscow, 119991 Russia
[b]Space Research Institute, Russian Academy of Sciences, Moscow, 117997 Russia
e-mail: siipatov@hotmail.com

**Abstract**—The dependences of inclinations of orbits of secondaries in the discovered trans-Neptunian binaries on the distance between the primary and the secondary, on the eccentricity of orbits of the secondary around the primary, on the ratio of diameters of the secondary and the primary, and on the elements of heliocentric orbits of these binaries are studied. These dependences are interpreted using the model of formation of a satellite system in a collision of two rarefied condensations composed of dust and/or objects less than 1 m in diameter. It is assumed in this model that a satellite system forms in the process of compression of a condensation produced in such a collision. The model of formation of a satellite system in a collision of two condensations agrees with the results of observations: according to observational data, approximately 40% of trans-Neptunian binaries have a negative angular momentum relative to their centers of mass.

## INTRODUCTION

The model of formation of solid planetesimals in the process of compression of rarefied condensations composed of dust and/or objects less than 1 m in diameter (i.e., boulder-sized objects) has received a great deal of attention lately (see, for example, Makalkin and Ziglina, 2004; Marov et al., 2013; Ziglina and Makalkin, 2016; Johansen et al., 2007; 2009a; 2009b; 2011; 2012; 2015a; 2015b; Cuzzi et al., 2008; 2010; Lyra et al., 2008; 2009; Chambers, 2010; Chiang and Youdin, 2010; Rein et al., 2010; Youdin, 2011; Youdin and Kenyon, 2013). Such condensations are also called preplanetesimals. The formation of planetesimals from rarefied condensations has been discussed since the 1950s (e.g., Safronov, 1972; Goldreich and Ward, 1973). New mechanisms of formation of condensations were proposed in recent years. The studies mentioned above and other studies dealing with the subject of the present paper were reviewed by Ipatov (2017).

In a radical departure from models of formation of binary small bodies at the stage of solid bodies (Goldreich et al., 2002; Weidenschilling, 2002; Funato et al., 2004; Astakhov et al., 2005; Richardson and Walsh, 2006; Ćuk, 2007; Pravec et al., 2007; Gorkavyi, 2008; Noll et al., 2008; Petit et al., 2008; Walsh et al., 2008), Ipatov (2009; 2010; 2014; 2017) and (Nesvorny et al., (2010)) have assumed that the majority of large trans-Neptunian binaries had formed as a result of compression of rarefied preplanetesimals. It was demonstrated in (Ipatov, 2010) that the angular momentum of two colliding rarefied preplanetesimals (relative to their center of mass), which moved in circular heliocentric orbits prior to the collision, may correspond to the angular momenta of known trans-Neptunian objects and asteroids with satellites (if the mass of a trans-Neptunian object or an asteroid is the same as the total mass of colliding preplanetesimals). Ipatov (2010; 2014; 2015a; 2015b; 2017) has assumed that the condensation producing a small body with a satellite (or satellites) had acquired most of its angular momentum in a collision of two condensations forming this parental condensation. It was found in (Ipatov, 2014; 2015b; 2017) that the angular velocities of condensations used in (Nesvorny et al., 2010) as the initial data for modeling the compression of a rarefied preplanetesimal and the resulting formation of a trans-Neptunian satellite system could be acquired in collisions of preplanetesimals with their radii comparable to the corresponding Hill radii. For example, these angular velocities fall within the range of values typical for a parental preplanetesimal forming as a result of a merger of two colliding rarefied preplanetesimals that had moved in circular heliocentric orbits before this collision. Ipatov (2017) has noted that the angular momentum of colliding preplanetesimals relative to their common center of mass could be positive or negative (depending on their orbits). The initial angular momenta of preplanetesimals were positive (Safronov, 1972). The overall angular momentum of a preplanetesimal collided with many small objects moved in weakly eccentric heliocentric orbits is also positive (Ipatov, 2000).

The observational data on binary trans-Neptunian objects are analyzed below. Specifically, the dependences of inclinations of orbits of secondaries in the discovered binaries on the distance between the primary and the secondary, on the eccentricity of orbits of the secondary around the primary, on the ratio of diameters of the secondary and the primary, and on the elements of heliocentric orbits of these binaries are studied. These dependences are interpreted using the model of formation of a satellite system in a collision of two rarefied condensations. In what follows, a binary object is understood as a satellite system with an arbitrary number of satellites. The two largest objects in such a system are called the primary and the secondary.

## PROGRADE AND RETROGRADE ROTATION OF TRANS-NEPTUNIAN BINARIES

Inclinations $i_s$ of orbits of secondaries around the primaries in 32 binary objects found in the trans-Neptunian belt (Fig. 1) and the origin of these inclinations are discussed in the present section. Figures 1–3 were plotted by browsing through the designations of trans-Neptunian satellite systems given in http://www.johnstonsarchive.net/astro/tnoslist.html and examining the data from http://www.johnstonsarchive.net/astro/astmoons/amNNN.html, where NNN is a unique combination of digits and letters corresponding to the designation of a certain satellite system. The data on this site are aggregated from multiple sources (see the references there). In the present study, only the data on satellite systems with known $i_s$ values were used ($i_s$ remains unknown for the majority of objects). Figures 1–3 present these data corresponding to the following 32 systems: 134340 (Pluto), 26308, 42355, 50000, 58534, 65489, 66652, 79360, 88611, 90482, 123509, 134860, 136108, 136199, 148780, 275809, 341520, 364171, 385446, 1998 WW31, 1999 OJ4, 2000 CF105, 2000 QL251, 2001 QW322, 2001 XR254, 2003 QY90, 2003 TJ58, 2003 UN284, 2004 PB108, 2005 EO304, 2006 BR284, and 2006 CH69. Only the largest satellite was considered in systems with two or more satellites. Note that $i_s$ is taken relative to the ecliptic and differs from the inclination relative to the plane perpendicular to the axis of rotation of the primary in the considered satellite system. For example, $i_s = 96°$ for Pluto, although Charon moves in the plane perpendicular to the rotational axis of Pluto. Since the orbits of secondaries are revised as more observational data become available, the orbital elements known to the reader of this paper may differ somewhat from the values given in Figs. 1–3.

*The fraction of secondaries with $i_s > 90°$ in the considered set of satellite systems is 12/32 = 0.375 at all values of eccentricity e of the heliocentric orbit of a binary and 12/28 ≈ 0.43 at e < 0.3*. If the binary with $i_s = 90.2°$ is excluded, the corresponding ratios for the remaining 31 binaries are 11/31 ≈ 0.355 and 11/27 ≈ 0.407. All four satellite systems with $e > 0.3$ have $i_s < 90°$.

The values of $i_s$ are distributed in a wide interval from almost 0° to 180° (Fig. 1a). This suggests that a considerable fraction of the angular momentum of the preplanetesimals that contracted to form trans-Neptunian satellite systems is not associated with the initial rotation of preplanetesimals or their collisions with small objects (e.g., dust and objects less than 1 m in diameter) and was acquired in collisions between preplanetesimals of similar masses, since the angular momentum would be always positive in the contrary case.

Ipatov (2010; 2017) has noted that the angular momentum of colliding preplanetesimals may be positive or negative depending on the heliocentric orbits of these preplanetesimals. If we consider the motion of preplanetesimals in unperturbed circular heliocentric orbits, the angular momentum of colliding preplanetesimals relative to their center of mass is negative for 20% of semimajor axes of the orbits. The fraction of collisions between preplanetesimals with negative angular momenta may exceed 20% if we consider the mutual gravitational influence of preplanetesimals and the fact that they could be smaller than Hill spheres and could have eccentric heliocentric orbits.

Kozlov and Eneev (1977) have calculated the dependence of the angular momentum of two material points relative to their center of mass on the distance between them. These material points moved around the Sun. Their heliocentric orbits were circular when the distance between the material points was equal to several Hill radii. Stating from this distance, the authors have integrated the three-body problem equations. The figure characterizing these calculations and presented in (Eneev and Kozlov, 1981; 2016) demonstrates that angular momentum $K_s$ of colliding preplanetesimals is negative when distance $d_c$ between their centers of mass at the moment of collision does not exceed radius $r_H$ of the Hill sphere for the sum of their masses, and difference $d_a$ between the semimajor axes of their heliocentric orbits, which were circular at a separation distance greater than a few Hill radii, falls between $2r_H$ and $2.9r_H$. In order to obtain distances expressed in Hill radii, one should multiply the distances given in the figure by $6^{1/3} \approx 1.82$. It follows from the analysis of this figure that $K_s > 0$ at $d_c = r_H$ and $0.45r_H < d_a < 2r_H$ (at $d_a < 0.45r_H$, no approaches to within $r_H$ are observed; i.e., $d_c > r_H$). Thus, the fraction of $d_a$ values corresponding to $K_s < 0$ is approximately 0.4 (≈0.9/(0.9 + 1.55)) at $d_c = r_H$. At lower $d_c$ values,

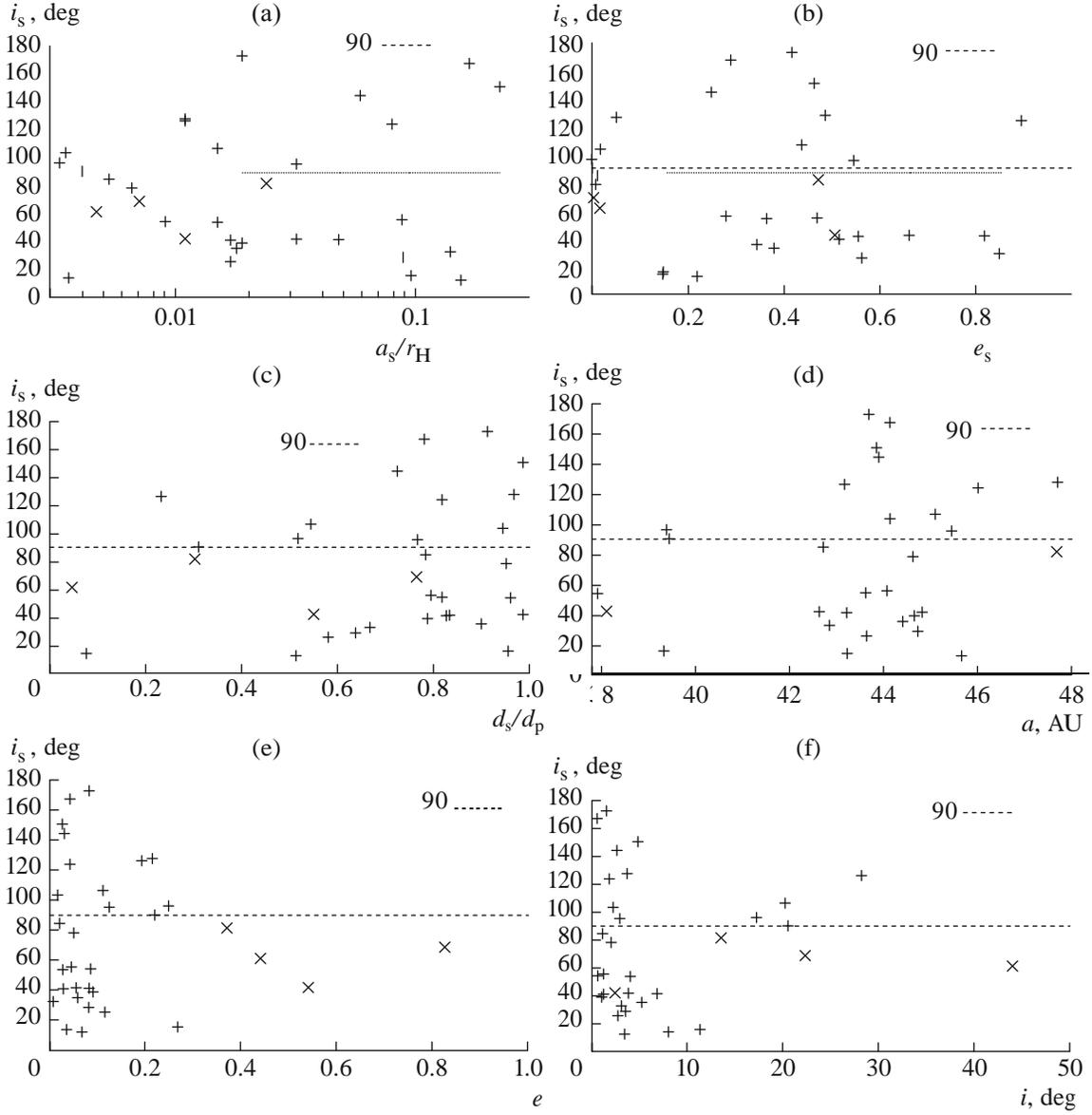

**Fig. 1.** Dependences of inclination $i_s$ of the orbit of the secondary around the primary of a trans-Neptunian binary on (a) ratio $a_s/r_H$ of distance $a_s$ between the secondary and the primary to radius $r_H$ of the Hill sphere of a binary system; (b) eccentricity $e_s$ of the orbit of the secondary around the primary; (c) ratio $d_s/d_p$ of diameters of the secondary and the primary; (d) semimajor axis $a$ of the heliocentric orbit of a binary; (e) eccentricity $e$ of the heliocentric orbit of a binary; (f) inclination $i$ of the heliocentric orbit of a binary.

The data for objects with $e < 0.3$ are denoted with plus signs (+), and the data for $e > 0.3$ are denoted with multiplication signs (×). The designations of the considered trans-Neptunian objects are given in the first section of the paper.

this fraction is closer to 0.5. Giuli (1968) has found that bodies colliding with the Earth may move in circular heliocentric orbits with $d_a \approx 2r_H$ prior to the collision. Ipatov (2000) has demonstrated that the fraction of collisions between celestial objects with highly eccentric heliocentric orbits and $K_s > 0$ differs by no more than a few percent from the fraction of collisions of such bodies with $K_s < 0$.

The number of observed binary trans-Neptunian objects with a positive angular momentum exceeds somewhat the number of discovered trans-Neptunian binaries with a negative angular momentum. This excess may be attributed in particular to the contribution of the initial positive angular momentum and the contribution of collisions between a preplanetesimal and small objects to the angular momentum of the parental preplanetesimal that produced the trans-Neptunian binary. The above estimates also demonstrate that *a certain excess of positive angular momentum in mutual collisions of preplanetesimals of similar sizes is possible.* If most of the angular momentum of the parental preplanetesimal compressing to form a binary was associated with the prograde rotation of the initial preplanetesimals or collisions between preplanetesimals and small objects, the angular momentum of the formed binary would be positive and $i_s$ would be close to zero.

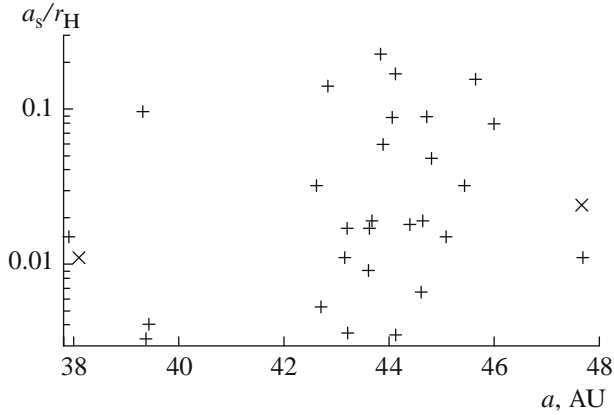 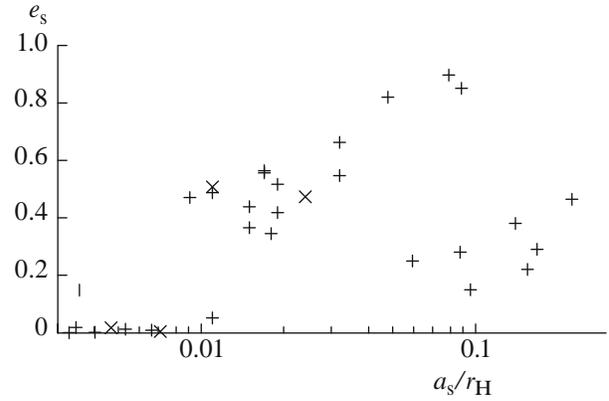

**Fig. 2.** Dependence of ratio $a_s/r_H$ of distance $a_s$ between the secondary and the primary to radius $r_H$ of the Hill sphere of a binary system on semimajor axis $a$ of the heliocentric orbit of a trans-Neptunian binary. The data for objects with eccentricity $e < 0.3$ of the heliocentric orbit of a binary are denoted with plus signs (+), and the data for $e > 0.3$ are denoted with multiplication signs (×).

**Fig. 3.** Dependence of eccentricity $e_s$ of the orbit of the secondary around the primary of a trans-Neptunian binary on ratio $a_s/r_H$ of distance $a_s$ between the secondary and the primary to radius $r_H$ of the Hill sphere of a binary system. The data for objects with eccentricity $e < 0.3$ of the heliocentric orbit of a binary are denoted with plus signs (+), and the data for $e > 0.3$ are denoted with multiplication signs (×).

Ipatov (2017) has compared the angular momentum of the parental preplanetesimal (condensation) needed to form a satellite system with the angular momentum obtained in a collision of two preplanetesimals. It was demonstrated that the contribution of a typical collision of identical preplanetesimals, which are similar in size to newly formed preplanetesimals, to the angular velocity of the parental preplanetesimal is several times higher than the angular velocity induced to the initial rotation of preplanetesimals. However, if the ratio of radii of colliding homogeneous preplanetesimals is higher than three or the radii of homogeneous preplanetesimals decreased by a factor of more than three in the interval from the moment of their formation (when they acquired the initial angular momentum) to collision, the contribution of the initial rotation of preplanetesimals to the angular momentum of the preplanetesimal formed in a collision is larger than that of the collision itself.

Based on these results published by Ipatov (2017), we can conclude that the observation of trans-Neptunian binaries with a negative angular momentum confirms that *certain homogeneous preplanetesimals have collided with homogeneous preplanetesimals of close sizes before their initial radii decreased by a factor of more than three*. This decrease in radii could be more significant for inhomogeneous (denser at the center) preplanetesimals. Ipatov (2017) has estimated the angular momentum of colliding preplanetesimals with circular heliocentric orbits. The results of calculations performed in (Ipatov, 2000) demonstrate that the fraction of collisions of preplanetesimals with a negative angular momentum and the angular momentum magnitude for eccentric heliocentric orbits may be higher than the corresponding values for circular heliocentric orbits. These results were discussed in more detail in (Ipatov, 1981a; 1981b). If the angular momentum values of colliding preplanetesimals in certain eccentric orbits are higher than the corresponding values in the case of circular heliocentric orbits, the ratio of radii of the initial preplanetesimals to the radii of colliding preplanetesimals that enables the formation of a satellite system may exceed the estimates obtained for circular heliocentric orbits.

## DISTANCES BETWEEN THE COMPONENTS OF A BINARY SYSTEM AT DIFFERENT HELIOCENTRIC ORBITS OF TRANS-NEPTUNIAN BINARIES

All four secondaries of the considered trans-Neptunian binaries with their heliocentric orbit eccentricity $e > 0.3$ move in prograde orbits around the primaries (i.e., $i_s < 90°$). At $e > 0.3$, ratio $a_s/r_H$ of distance $a_s$ between the primary and the secondary (semimajor axis $a_s$ of the orbit of the secondary around the primary) to Hill radius $r_H$ of a binary object is below 0.024, while $a_s/r_H$ at $e < 0.3$ may be as high as 0.225 (Fig. 1a). Note that trans-Neptunian objects with $e > 0.3$ could form in the feeding zone of giant planets (see, e.g., Ipatov, 1987; Levison and Stern, 2001; Gomes, 2003; 2009); in other words, they could form closer to the Sun than the objects with $e < 0.3$. After the binaries had moved from this feeding zone to the trans-Neptunian belt and their semimajor axes (and $r_H$ values) had increased, the values of $a_s/r_H$ became smaller. For example, semimajor axis $a$ of the orbit of the 65 489 Ceto-Phorcys binary is 102 AU; therefore, ratio $a_s/r_H$ for this object could decrease by a factor of five in the process of its migration from the place of origin to the current orbit.

Figure 2 shows that *the average $a_s/r_H$ values at $42 < a < 46$ AU are higher than at $38 < a < 40$ AU*, where $a$ is the semimajor axis of the heliocentric orbit of a binary. This difference may be attributed to the fact that the average ratios of radii of collided preplanetesimals to their Hill radii are lower at shorter distances between preplanetesimals and the Sun at $38 < a < 46$ AU. These ratios could be lower, for example, due to the faster contraction of preplanetesimals at shorter distances from the Sun. It is also possible that the objects with $38 < a < 40$ AU normally formed with such semimajor axes $a$ that are shorter than their current values. The $a_s/r_H$ ratio decreased as these objects moved further away from the Sun.

## RELATIONSHIPS BETWEEN THE ELEMENTS OF ORBITS OF SECONDARIES OF TRANS-NEPTUNIAN BINARIES

*At $a_s/r_H < 0.008$, where $a_s$ is the distance between the primary and the secondary and $r_H$ is the Hill radius, the values of inclination $i_s$ of the orbit of the secondary around the primary are* (except for one object) *distributed between 60° and 105°; i.e., they are in the neighborhood of 90°* (Fig. 1a). This may probably be attributed to the fact that the ratios of radii of binary-forming preplanetesimals to $r_H$ (and, consequently, the ratios of these radii to the thickness of the disk within which these preplanetesimals moved) could be lower at smaller $a_s/r_H$ values. If this was the case, instead of moving in approximately the same plane (as expected for preplanetesimals with their diameters being roughly equal to the disk thickness), colliding preplanetesimals often moved one above another. Note that the semimajor axes and eccentricities of heliocentric orbits of trans-Neptunian binaries with $a_s/r_H < 0.008$ vary widely: from 39 to 102 AU and from 0.02 to 0.8, respectively. Therefore, such objects could form at various distances from the Sun.

Eccentricities $e_s$ of orbits of secondaries around the primaries of trans-Neptunian binaries were below 0.15 at $a_s/r_H < 0.008$ (Fig. 3). *Thus, orbits close to the primary have low eccentricities.* The objects with $e_s < 0.1$ have $a_s/r_H < 0.0011$. Eccentricities $e_s$ are normally above 0.2 at $a_s/r_H > 0.011$ (Fig. 3). The values of $e_s$ typically fall within the range of 0.3–0.7 at $0.009 < a_s/r_H < 0.035$ and vary in a wider range of 0.15–0.9 at $a_s/r_H > 0.035$. *The correspondence between higher maximum values of $e_s$ and higher values of $a_s/r_H$ fits the model of formation of satellites from a disk of material* (the disk may form, for example, as a result of contraction of a rarefied condensation). The formation of planetary satellites from a disk is a widely used model (see, e.g., Ruskol, 1982; Vityazev et al., 1990). The orbits of planetary satellites are also almost circular at small distances from planets and may be eccentric at greater distances.

Figure 1b demonstrates that inclinations $i_s$ of orbits of secondaries of trans-Neptunian binaries fall within the range of 60°–130° at eccentricities $e_s$ below 0.1, but may assume arbitrary values at higher $e_s$. The $a_s/r_H$ ratio for objects with $e_s < 0.1$ did not exceed 0.008. It was already noted above that colliding preplanetesimals could move one above another in this case. This resulted in a considerable inclination of the angular momentum vector of the parental preplanetesimal with respect to the ecliptic.

## INCLINATIONS OF ORBITS OF SECONDARIES OF TRANS-NEPTUNIAN BINARY OBJECTS AT VARIOUS RATIOS OF DIAMETERS OF COMPONENTS OF THESE OBJECTS

Inclination $i_s$ of the orbit of the secondary around the primary of a binary trans-Neptunian object may assume arbitrary values at ratios $d_s/d_p > 0.7$ of diameters of the secondary and the primary. However, no binaries with $130° < i_s < 180°$ and $d_s/d_p < 0.7$ were found, and only a single object with $i_s < 50°$ and $d_s/d_p < 0.5$ was observed (Fig. 1c). *The lack of trans-Neptunian binaries with $i_s > 130°$ at $d_s/d_p < 0.7$* may be attributed to the fact that the contribution of the initial positive angular momentum of preplanetesimals to the angular momentum of the parental preplanetesimal, which had contracted and formed the binary trans-Neptunian object, was greater (and the angular momentum acquired in a collision of preplanetesimals of close masses, which produced the parental preplanetesimal, was lower) at $d_s/d_p < 0.7$ than at $d_s/d_p > 0.7$. The contribution of the angular momentum acquired in a collision could be lower at lower $d_s/d_p$ ratios since the masses of colliding preplanetesimals differed more in this case than at higher $d_s/d_p$ values or due to the higher compression of preplanetesimals before their collision. The fraction of discovered trans-Neptunian binaries with $d_s/d_p > 0.7$ is $20/32 \approx 0.625$. A significant (~0.8) fraction of trans-Neptunian binaries with $d_s/d_p > 0.7$ was also obtained in computer simulations performed by Nesvorny et al., 2010.

No dependence of $i_s$ on diameter $d_p$ of the primary was found. Diameters $d_s$ of secondaries in 32 chosen trans-Neptunian satellite systems varied from 50 to 1200 km; in approximately one half (47%) of these systems, $d_s < 100$ km. Diameters $d_p$ of the primary fell within the range of 65–2400 km.

# DEPENDENCES OF THE ORBIT INCLINATION OF THE SECONDARY OF A TRANS-NEPTUNIAN BINARY ON THE HELIOCENTRIC ORBITAL ELEMENTS OF THIS BINARY

Semimajor axes $a$ of heliocentric orbits of most trans-Neptunian binaries considered in the present study fall within the range of 42.5–46 AU. Their eccentricities $e$ are typically below 0.1. Three plutinos in the 2 : 3 resonance with Neptune and two objects in the 1 : 2 resonance (twotinos) are also present in the sample. Their eccentricities exceed 0.2. The eccentricities of three objects with semimajor axis $a$ = 38.1, 67.8, and 101.9 AU exceed 0.4. Large eccentricities are the result of evolution of orbits of trans-Neptunian objects. The objects with such eccentricities could be formed in another region (closer to the Sun).

The dependences of inclination $i_s$ of the orbit of the secondary around the primary of a trans-Neptunian binary (or an object with several satellites) on the heliocentric orbital elements $(a, e, i)$ of this object are presented in Figs. 1d–1f. The values of $i_s$ are lower than 110° at $38 < a < 40$ AU. *At $38 < a < 44$ AU, the maximum $i_s$ values are higher at larger semimajor axes $a$ of the heliocentric orbit* (Fig. 1d). This dependence may be attributed to the fact that the contribution of positive angular momentum increments is larger at shorter distances from the Sun (due, e.g., to lower eccentricities of heliocentric orbits of colliding preplanetesimals and a more significant contribution from small objects falling onto preplanetesimals). It was noted above that the typical $a_s/r_H$ values at $38 < a < 44$ AU also increase with $a$.

The maximum value of $i_s$ is normally lower at higher eccentricities $e$ of the heliocentric orbit (Fig. 1e). It is close to 180° at $e < 0.1$, is roughly equal to 128° at $e \approx 0.2$, and is below 90° at $e > 0.3$. Such $i_s$ values at $e > 0.3$ suggest that the contribution of the initial positive angular momentum of preplanetesimals and/or the contribution of small objects falling onto preplanetesimals to the resulting angular momentum of the parental preplanetesimal in the region of formation of trans-Neptunian binaries with $e > 0.3$ could be larger than that for other trans-Neptunian objects. The objects with $e > 0.3$ could form not in the trans-Neptunian belt, but in the feeding zone of giant planets (closer to the Sun). The large contribution of small objects to the final angular momentum of the parental preplanetesimal that produced a trans-Neptunian object with $e > 0.3$ may be related to the fact that the fraction of small objects in the feeding zone of giant planets was higher than that in the trans-Neptunian belt.

At inclinations $i > 13°$ of the heliocentric orbits of trans-Neptunian binaries and their eccentricities $e \geq 0.219$, the values of $i_s$ are within a certain neighborhood of 90° ($61° \leq i_s \leq 126°$, Fig. 1f); specifically, $68° < i_s < 110°$ at $13° < i < 24°$. It is likely that trans-Neptunian binaries with $i > 13°$ formed when the disk thickness exceeded the sizes of colliding preplanetesimals, and these colliding preplanetesimals could go one above another. The chances of obtaining both an $i_s$ value close to 90° and a large $i$ value are higher in this case.

The relationships between elements of the orbit of the secondary around the primary of a binary system, elements of the heliocentric orbit of a binary system, and diameters of the secondary and the primary were interpreted in the present study. This interpretation demonstrates that *the proposed model agrees with the results of observations of trans-Neptunian binaries. According to this model, a satellite system forms as a result of compression of a parental preplanetesimal that acquired a considerable fraction of its angular momentum in a collision of two preplanetesimals*. Any other theory of formation of trans-Neptunian binaries must explain the observations presented in Figs. 1–3 (e.g., must explain why a considerable fraction of trans-Neptunian binaries have negative angular momenta).

## CONCLUSIONS

The dependences of inclinations of orbits of secondaries in trans-Neptunian binaries on the semimajor axis and the eccentricity of orbits of the secondary around the primary, on the ratio of diameters of the secondary and the primary, and on the elements of heliocentric orbits of these binaries were studied by analyzing observational data. These dependences were interpreted for the first time within the model of formation of a satellite system from a rarefied condensation composed of dust and/or objects less than 1 m in diameter. It is assumed in this model that the parental condensation was formed and acquired a considerable fraction of its angular momentum in a collision of two condensations. Approximately 40% of trans-Neptunian binaries have a negative angular momentum relative to their centers of mass. The model of formation of a satellite system in a collision of two condensations explains these observational data.

## ACKNOWLEDGMENTS

The author would like to thank the referees for helpful remarks.

This study was supported in part by the Russian Foundation for Basic Research (projects nos. 14-02-00319 and 17-02-00507 A) and Program no. 7 of the Presidium of the Russian Academy of Sciences.